\documentstyle[12pt,aps,preprint]{revtex}
\input epsf
\begin{document}
\tightenlines

\title{Quantum percolation in power-law diluted chains}

\author{R. P. A.  Lima and M. L. Lyra}
\address{Departamento de F\'{\i}sica,
Universidade Federal de Alagoas, 57072-970 Macei\'o - AL, Brazil}
\maketitle

~

~

\begin{abstract}

We investigate the quantum percolation problem in a diluted
chain with long-range hopping amplitudes. Each bond is
activated with  probability $p(r) = p_1/r^{\alpha}$, where $r$ is the distance between two
sites and $\alpha$ characterizes the range of the interactions.  The average
 participation ratio
of all eigenstates is used as a measure of the wave-functions localization
length.
We found that, above a quantum percolation threshold $p_1^{(q)}$,
 true extended states appears for
 $\alpha < 1.5$. In the
regime of $1.5 < \alpha <2.0$ there is no trully extended states even in the
presence of a spanning cluster. Instead, a phase of critical
wave-functions sets up.
\end{abstract}

\newpage

The quantum percolation phenomenon has been extensively investigated during
the last decade. Besides its own challenging theoretical aspects, quantum percolation
has also an appealing technological interest
 due to its possible relevance for the underlying mechanism
leading to high-$T_c$ superconductivity as well as to its potential application
for the development of nano-structure devices\cite{mookerjee95,phillips91,dallacasa95,kodama99}.
 In these systems, the one-electron
eigenstates on a site or bond diluted lattice can exhibit a localization/delocalization
transition as a function of the disorder strength in a close relation to the Anderson
transition\cite{anderson58}. At low dimensions $d \leq 2$ the eigenstates are believed to be
 localized for any degree
of disorder even in the presence of spanning clusters of connected sites.

Several approaches  have been
employed for studying the nature of one-electron eigenstates in diluted lattices.
The most usual numerical algorithms  put forward Green's function, renormalization group,
transfer matrix or diagonalization
techniques, or a combination of these, to explicitly compute some relevant quantities
such as the wave-function localization length, reflection and transmission coefficients,
conductance and spectral statistics\cite{mookerjee95,chang87,root88,soukoulis92,berkovits96,kaneko99}.
 The results obtained from numerical works
on finite lattices have to be
extrapolated by exploring the finite size scaling hypothesis to
get the true thermodynamic behavior. In spite of all efforts made to fully
understand all features related to the quantum percolation phenomenon, some relevant
points are still open to debate. In $3D$ lattices, estimates from different techniques
of the critical
exponent $\nu$ describing the divergence of the localization length are still quite
scattered and often differ from that expected for the Anderson model\cite{chang87,root88,soukoulis92,berkovits96}.
 However,
recent energy level statistics
analysis have reported results supporting that quantum percolation indeed belongs to the
Anderson transition universality class\cite{kaneko99}. In $2D$ the present picture is also not firmly settled.
Although all states are believed to be localized according to  Anderson-like
scaling   arguments, some studies have reported signatures of a localization-delocalization
transition\cite{meir89,koslowski90,daboul00}.

In the present work, we introduce a simple but rather non-trivial model which
exhibits a quantum percolation transition in one-dimension. We
consider a bond diluted tight-binding Hamiltonian with long-range hopping amplitudes
which are activated with probability $p(r_{i,j})=p_1/r_{i,j}^{\alpha}$, where
$r_{i,j}$ is the distance between sites $i$ and $j$. Models with long-range
interactions\cite{mirlin96,mirlin00,parshin98,cressoni98} as well as with
long-range correlated disorder\cite{moura98,izrailev99}
have already been used in the context of the Anderson transition. In these models a
localization/delocalization transition have been identified even at low-dimensions.

In the classical
counterpart of the long-range diluted model, a percolating cluster emerges above a
critical value $p_1^{(c)}$ provided that $\alpha \leq 2$\cite{schulman83,rego99}. For $\alpha
\leq 1$, a percolating  cluster is always present for any finite
$p_1$. At $\alpha=2$ the critical concentration is finite and its
value was recently conjectured to be $p_1^{(c)} = 0.703...$\cite{rego99,cannas97}.
This later particular case has attracted much interest once it
depicts
an intermediate phase with slowly decaying correlations\cite{imbrie88}.
In the range  $1.5 < \alpha < 2$, the critical exponents of the Ising version of this model
are non-trivial and change continuously with $\alpha$\cite{barati00}. For  $1.0< \alpha < 1.5$, the transition
is
characterized by
classical mean-field
exponents\cite{luijten97,cannas95}. Here, we will
 show some results from exact diagonalization combined with
finite size scaling for the average participation ratio of all
one-electron eigenstates of the diluted long-range tight-binding model.
 These indicate that this model has also a rich
phenomenology concerning the quantum percolation transition.

The quantum percolation problem in a
 bond diluted chain with non-random long-range hopping amplitudes
can be represented by a one-electron tight-binding Hamiltonian
with a single orbital per site. In the atomic orbital wave-function
basis $\{|n\rangle \}$, it is expressed as
\begin{equation}
{\cal H}=\sum_n\epsilon_n|n\rangle \langle n| + t\sum_{n\neq m} h(r=|n-m|)[|n\rangle\langle m|]~~~,
\end{equation}
where the sum extends over the $L$ sites of a chain with open
boundaries. Hereafter, we will use energy units such that the
hopping amplitude $t=1$. To avoid degeneracies during the
numerical diagonalization procedure, the orbital energies $\epsilon_n$ will be considered
to have a very weak disorder being uniformly distributed in the
interval $[-W,W]$ with $W=0.0005$. $h(r)$  are binary
distributed following
\begin{equation}
{\cal P}[h(r)] = p(r)\delta[h(r)-1]+ [1-p(r)]\delta[h(r)]~~~,
\end{equation}
with power-law decaying probabilities $p(r)=p_1/r^{\alpha}$, where
the probability of a first-neighbors bond being activated $p_1
\leq 1$ and $\alpha$ governs the long-range character of the interactions.
Note that even in the case of $p_1=1$ disorder is present
through the dilution of further neighbors bonds and the
wave-functions may become localized if these probabilities decay
fast enough. On the other hand, for slowly decaying probabilities,
the high connectivity of the chain may stabilize
extended states.

In what follows we diagonalize the Hamiltonian (1) in finite chains of size up to $L=1,600$
to obtain all normalized
one-electron eigenfunctions $\Psi^{(k)} = \sum_n
\psi_n^{(k)}|n\rangle$. Distinct realizations
of the disorder were implemented such that for each chain size
we computed $32,000$ eigenfunctions. The
localized/delocalized nature of the wave-functions is
characterized by the behavior of the participation ratio $P^{(k)} =
1/ \sum_n [\psi_n^{(k)}]^4$. For exponentially localized states, the
participation ratio gives an estimate for the localization length
$\xi$. For truly extended states it scales as $L$. At the
localization/delocalization transition it depicts a power-law
growth $P\propto L^{\tau_2}$ with $\tau_2 <1$ due to the power-law tails and
the multifractal character of the
critical wave-functions\cite{schreiber91,janssen98,kravtsov97,nishigaki99}.

The density of states for this model can be directly measured from
the distribution of energy eigenvalues. Typical integrated density
of states (IDOS) are shown in figure~1 as obtained from  a lattice of
$L=1600$ and $\alpha = 1.25$.  For $p_1=0.1$ there is no spanning cluster\cite{rego99}.
The  density of states gaps and the delta-function singularities
reflect the chain's finite clusters structure. All
wave-functions are then localized within the finite clusters.
In the opposite situation of high bond concentration ($p_1=0.9$), previous Monte Carlo
simulations\cite{rego99}
have shown that a spanning cluster is present. In this situation, the density of
state turns out to be gapless.

In order to investigate the possible existence of delocalized
states in the regime where a spanning cluster is present, we computed the average
participation ratio of the wave-functions with energy between $E$ and $E+\Delta E$ as a function of $E$
for distinct chain sizes.
The interval $\Delta E$ was chosen to be much smaller than the band width but large enough to contain many
states. Figure 2a
shows the result corresponding to localized states when only
finite clusters are present ($p_1=0.1$, $\alpha =1.25$). The participation ratio is
$L$-independent as expected. For the same value of $\alpha$ but with a
higher concentration $p_1=0.9$,
we obtained the participation ratio to scale as $P(E)\propto L$
within the entire band. This corroborates our preliminary idea that the presence of long-range
bonds with slow-decaying probabilities favors the emergence of delocalized states.

To locate the critical concentration $p_1^{(q)}$ above
which extended states appear, we computed the average participation
ratio of all eigenstates within the band $P=\langle P^{(k)}\rangle_k$ as a function of the
first neighbors concentration $p_1$. The results for $\alpha =
1.25$ are shown in figure~3. It clearly depicts a transition from
localized to delocalized states with the critical quantum percolation threshold
being roughly  $p_1^{(q)}=0.25(3)$. This value
is above, although relatively close to, a recent Monte Carlo estimate for the classical
percolation threshold\cite{rego99} such that the general trend $p^{(q)} >p^{(c)}$
seems also to hold for the present model. However, the error bars on both
estimates of $p_1^{(c)}$ and $p_1^{(q)}$ are still large to firmly state this
trend.
Within the range $1.0<\alpha<1.5$, the above picture remains
with the critical concentration monotonically increasing with
$\alpha$.

In the regime of $1.5 < \alpha < 2.0$, our data for the
participation ratio exhibit only a slow growth with increasing chain size.
In figure~4, we show the
results for $\alpha
= 1.75$. In this case, we have a transition from localized states
to critical states having $P\propto L^{\tau_2}$, with $\tau_2<1$ (see inset).
Contrary to the classical percolation case, there is no
trully extended one-electron states in this regime. The estimated critical
concentration separating localized and critical states $p_1^{(q)}(\alpha = 1.75) = 0.60(5)$
is also
relatively above
the numerical estimate of classical percolation threshold\cite{rego99}. In figure~5, we plot two
typical wave-functions representing an extended and a critical
state. Notice that the critical wave-function exhibits spikes of
many sizes reflecting its multifractal nature\cite{schreiber91,janssen98,kravtsov97,nishigaki99}.

In summary, we studied the quantum percolation problem in a
one-dimensional tight-binding Hamiltonian with diluted long-range
hopping amplitudes whose probability of occurrence decay as $p(r) =
p_1/r^{\alpha}$. For the regime of $1.0<\alpha<2.0$, a spanning
cluster of connected sites emerges above a classical percolation
threshold $p_1^{(c)}$ with mean-field like exponents for $\alpha <
1.5$ and non-trivial ones for $\alpha > 1.5$.
Using  numerical diagonalization on finite lattices and finite size
scaling,
we found that for $1.0< \alpha < 1.5$ the one-electron eigenstates  exhibit
a localization/delocalization transition with increasing first-neighbors bond concentration $p_1$.
The quantum critical percolation  was found to be above, although relatively
close to, the  classical threshold. In the regime of $1.5< \alpha
< 2.0$, no truly extended states appears even for $p_1=1$. Instead, a phase
transition between localized and critical states takes place.

The diagonalization procedure employed in the present work does not allow much larger
systems to be investigated within a reasonable computational
effort. Therefore, precise estimates of  quantum critical
thresholds and exponents governing the presently reported transitions
require the use of alternative methods that are able to overcome the large corrections
to scaling present in this model, specially at the vicinity of the $\alpha = 2$ anomaly.
Numerical techniques which are less sensitive to
geometrical problems, such as energy levels and multifractal
analysis, would provide further relevant insights on these points.
Also,  analytical approaches along the lines used to
investigate the power-law random band Anderson model\cite{mirlin96,mirlin00} would
certainly be valuable. We hope the present report can stimulate
further works in these directions.

This work was partially supported by the Alagoas state research agency FAPEAL and
the Brazilian agencies CNPq and CAPES.

\newpage

\section{Figure captions}

\noindent
{\bf Figure 1 -} The integrated density of states (IDOS) as a function of
$E/t$ for a linear chain with $L=1,600$ sites and
$\alpha = 1.25$. A configurational average over $20$ realizations of
the disorder was employed. For $p_1=0.1$, all clusters are finite.
The gaps and delta function singularities in the density of states
reflect the absence of a spanning cluster. For $p_1=0.9$, a
spanning cluster is present and the density of states becomes smooth.

\noindent
{\bf Figure 2 -} The participation ratio as a function $E/t$ for chains with $\alpha =
1.25$. (a) $p_1=0.1$ and linear sizes $L=800$ (circles) and $1,600$ (triangles):
all eigen-functions are
exponentially localized and the participation ratio is roughly size
independent; (b) $p_1=0.9$ and linear sizes $L=800$ (diamonds) and $1,600$ (squares):
all eigenfunctions are extended
and the participation ratio scales linearly with $L$.

\noindent
{\bf Figure 3 -} The average participation ratio (normalized by
the chain's size) as a function of the first-neighbors concentration
$p_1$ and distinct lattice sizes $L=200$ (circles), $400$ (squares),
$800$ (diamonds) and $1,600$ (triangles) for $\alpha = 1.25$.
It shows a
clear localization/delocalization transition around $p_1^{(q)} =
0.25(3)$. This behavior is typical for the range $1<\alpha<1.5$ with
an $\alpha$-dependent critical concentration.

\noindent
{\bf Figure 4 -}  The average participation ratio
as a function of the first-neighbors concentration
$p_1$ and distinct lattice sizes $L=200$ (circles), $400$ (squares),
$800$ (diamonds) and $1,600$ (triangles) for $\alpha = 1.75$.
It shows a crossover from a phase of localized states at low
concentrations to a phase of weakly delocalized states at
high concentrations. Inset: the size dependence of the
participation ratio for concentrations $p_1=0.2$ (diamonds), $0.6$ (triangles)
and $1.0$ (circles).
At high concentrations, a phase of critical states sets up
where the participation ratio scales as $P\propto L^{\tau_2}$. For
$p_1=1$ we obtained $\tau_2 = 0.73(2)$. At the estimated critical concentration
$p_1^{(q)}=0.60$, we found $\tau_2 = 0.38(2)$. This behavior is typical
for the range $1.5<\alpha<2.0$ with $\alpha$-dependent critical
concentration and exponents.

\noindent
{\bf Figure 5 -} Representative extended (bottom) and critical (top) wave-functions.
These were obtained
from a $L=1,600$ chain, $p_1=1.0$, $\alpha = 1.25$ (extended state) and
$\alpha = 1.75$ (critical state). The extended state presents a
random uniform distribution over all chain. The critical state has
spikes of many sizes characterizing its multifractal distribution.

\newpage

\begin{figure}
\setlength{\epsfxsize}{12.cm}
\centerline{\mbox{\epsffile{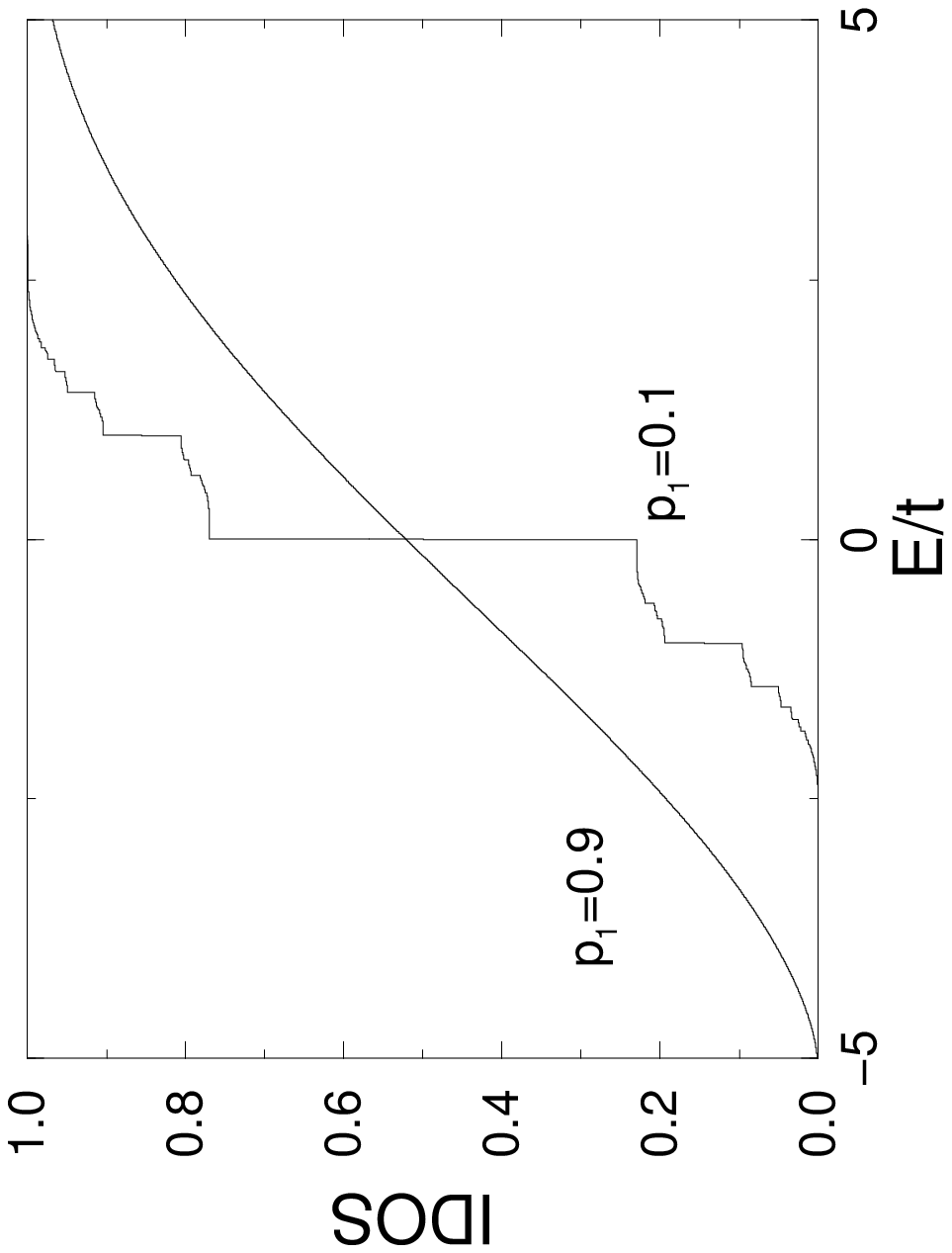}}}
\caption{~}
\label{fig1}
\end{figure}

\newpage

\begin{figure}
\setlength{\epsfxsize}{12.cm}
\centerline{\mbox{\epsffile{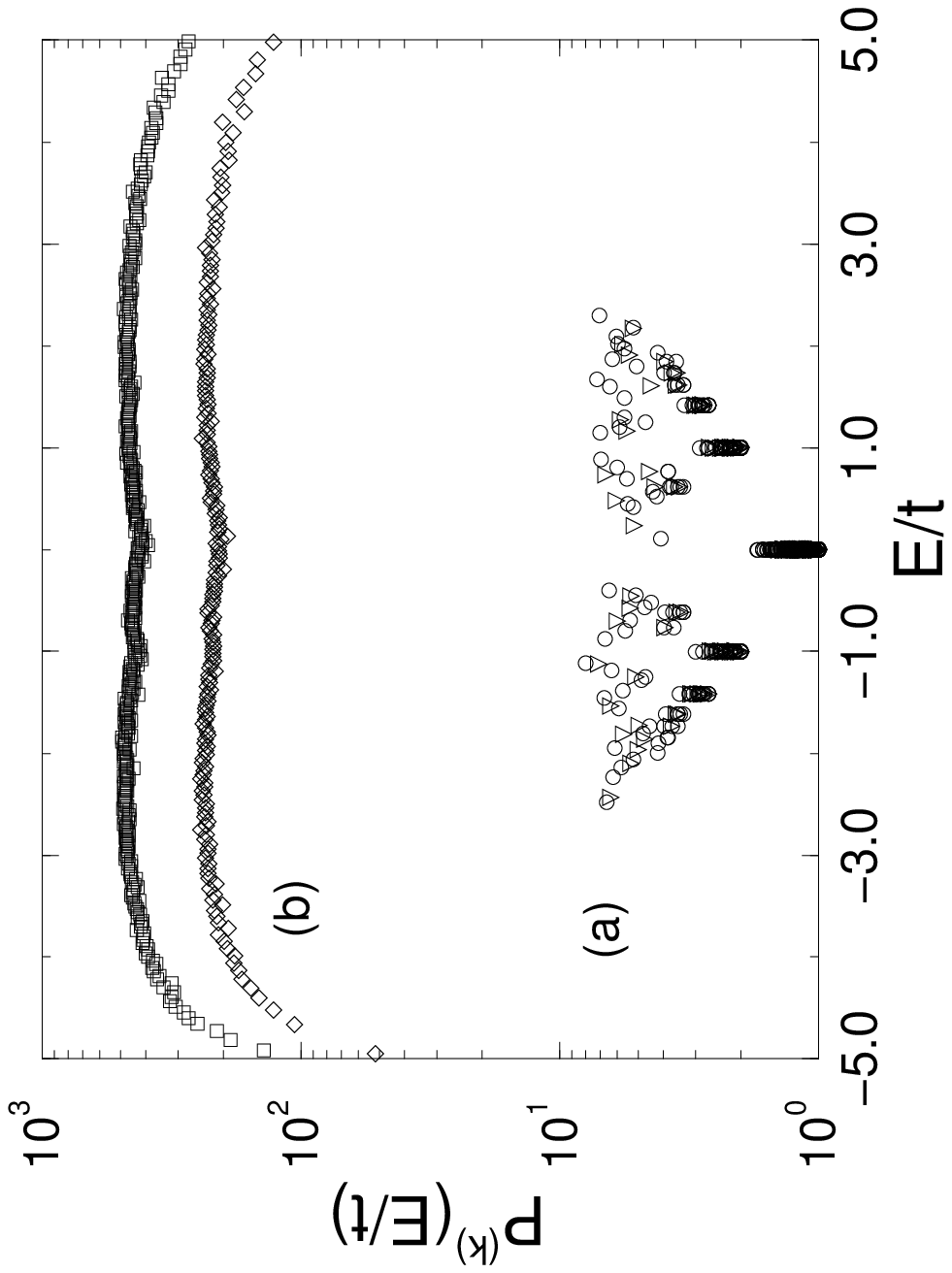}}}
\caption{~}
\label{fig2}
\end{figure}

\newpage

\begin{figure}
\setlength{\epsfxsize}{12.cm}
\centerline{\mbox{\epsffile{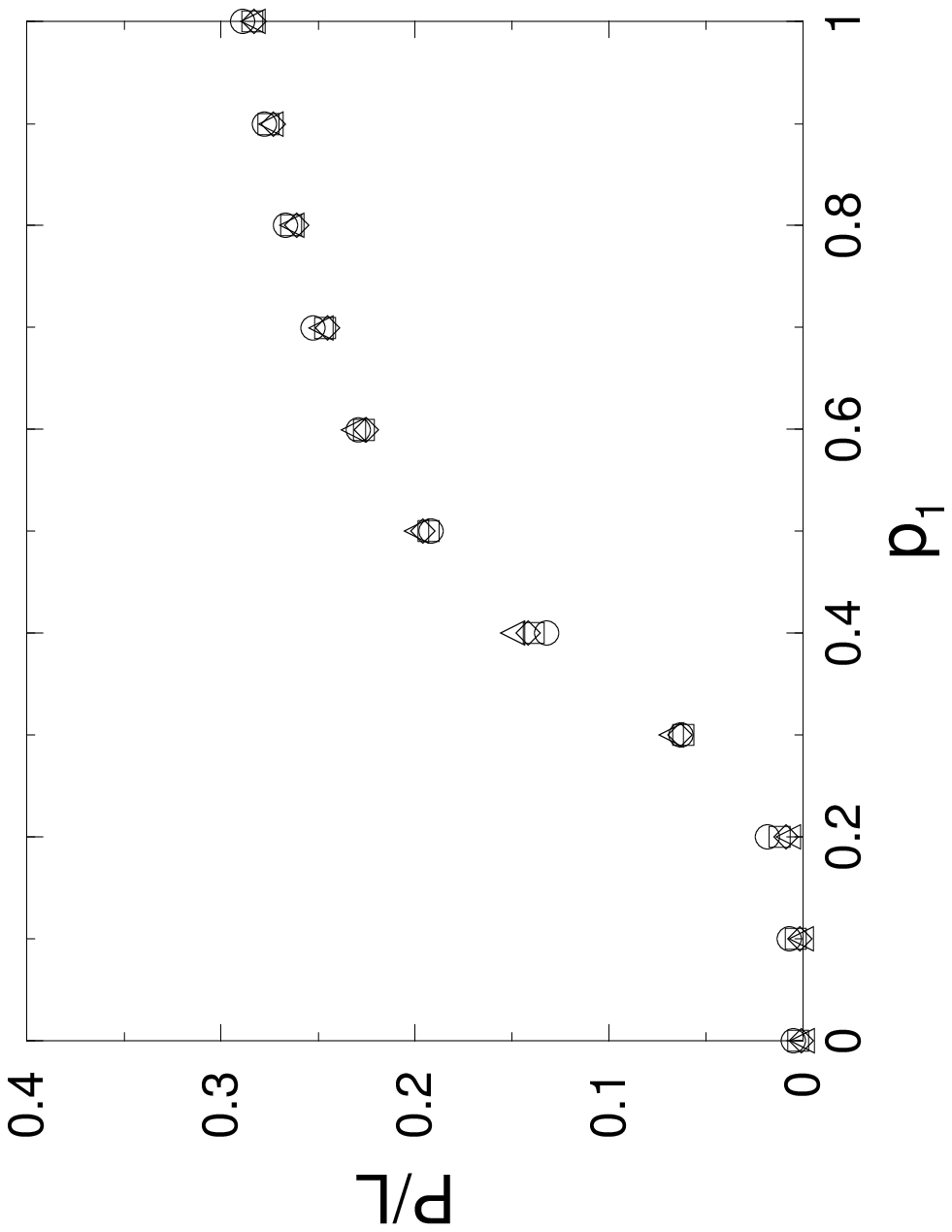}}}
\caption{~}
\label{fig3}
\end{figure}

\newpage

\begin{figure}
\setlength{\epsfxsize}{12.cm}
\centerline{\mbox{\epsffile{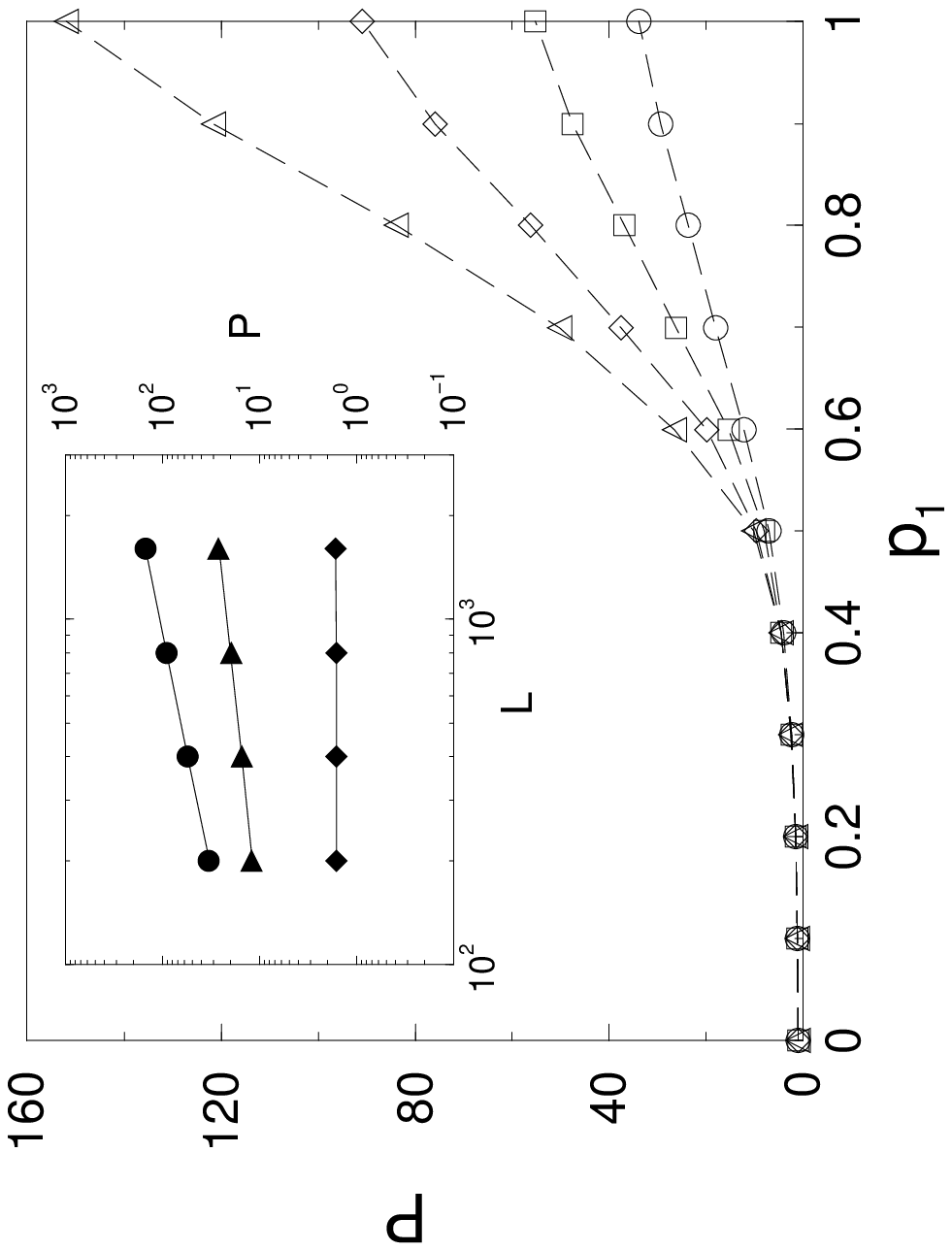}}}
\caption{~}
\label{fig4}
\end{figure}

\newpage

\begin{figure}
\setlength{\epsfxsize}{12.cm}
\centerline{\mbox{\epsffile{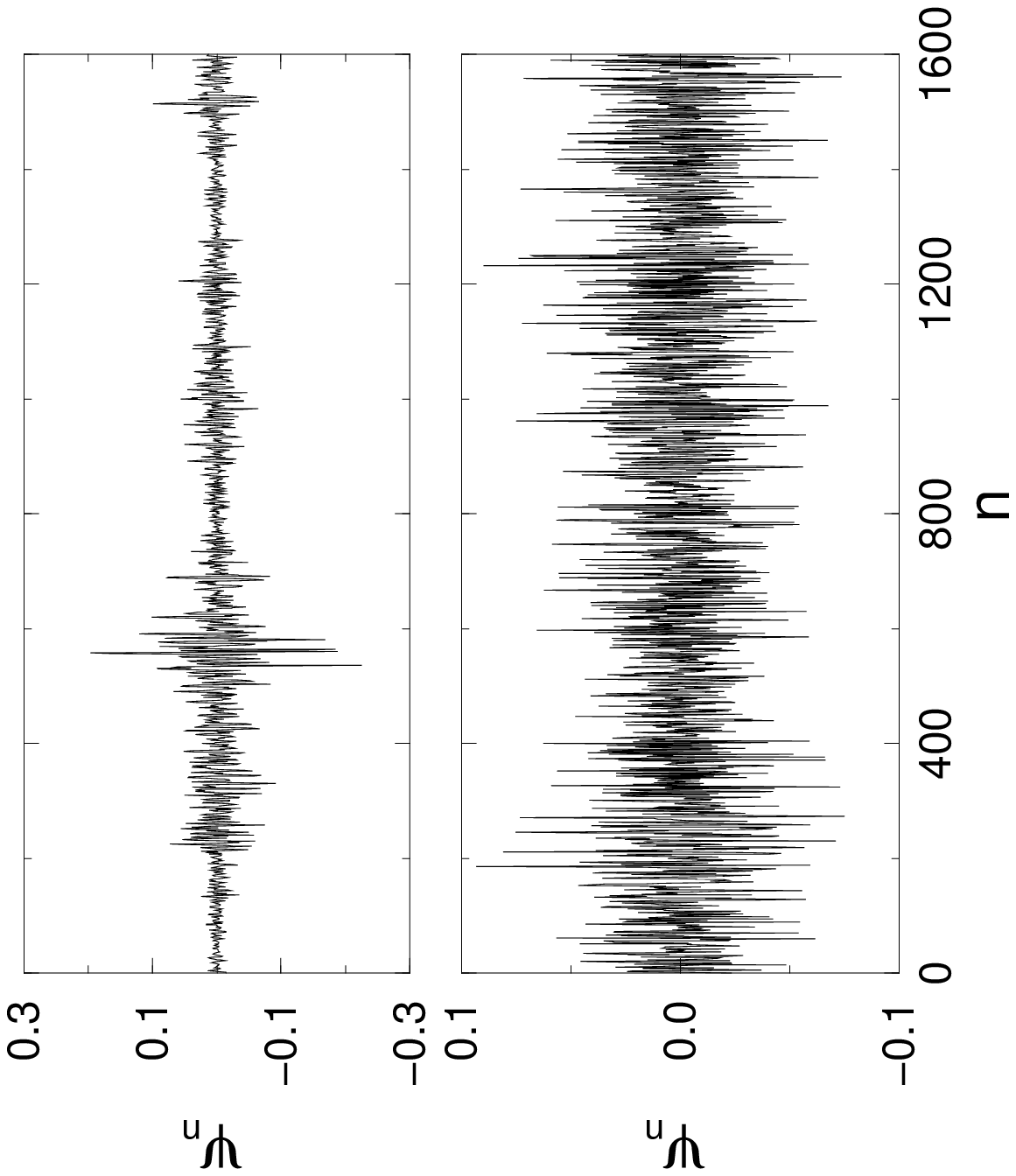}}}
\caption{~}
\label{fig5}
\end{figure}

\end{document}